\documentclass[11pt]{article}
\usepackage{moriond,epsfig}

\def\be{\begin{equation}}
\def\ee{\end{equation}}
\def\bea{\begin{eqnarray}}
\def\eea{\end{eqnarray}}

\begin{document}
\vspace*{4cm}
\title{Runaway electromagnetic cascade in shear flows and
high energy radiation of astrophysical jets}

\author{B.E. STERN $^{1,2,3}$, J. POUTANEN$^2$  }

\address{$^1$Astro Space Center of Lebedev Physical Institute,
Moscow, Profsoyuznaya 84/32, 117997, Russia\\
$^2$Astronomy Division, P.O.Box 3000, 90014 University of Oulu,
Finland\\
$^3$ Institute for Nuclear Research RAS, Moscow, Profsojuznaya 7a}

\maketitle\abstracts{
 We propose a straightforward and efficient mechanism of the high energy
emission of astrophysical jets associated with an exchange of interacting
high energy photons between the jet and external environment and vice versa.
Interactions
which play the main role in this mechanism, are $e^+ e^-$ pair production by
photons and inverse Compton scattering. The process has been studied with
numerical simulations demonstrating that under reasonable conditions it has a
supercritical character: high energy photons breed exponentially being fed
directly by the bulk kinetic energy of the jet. Eventually, there is a
feedback of particles on the fluid dynamics and the jet partially decelerates.
 }

\section{Introduction}

Traditionally, the dissipation of a relativistic bulk motion into radiation
is assumed to be associated with the shock acceleration of {\it charged}
particles. The latter should certainly play some role, however,
at least in a relativistic case their role is limited by a number of
factors\cite{ach01,bo01}. On the other hand, in some cases
(blazars or gamma-ray bursts) the energy release in gamma-rays
can be comparable to the estimated total kinetic energy of the bulk motion
in the source
and one has to search for a more efficient mechanism.

 Derishev et al.\cite{der03}   and Stern\cite{st03} independently suggested
that interacting {\it neutral} particles can convert bulk kinetic energy
into radiation more efficiently than this can be done by charge particles.
Indeed, they easily cross the shock front or the boundary of the shear layer
and can be converted into charge particles (e.g. via $e^+ e^-$ pair
productions by two photons) inside the relativistic fluid.
Then, secondary charged particles gain a factor $\sim \Gamma^2$ (where
$\Gamma$ is the Lorentz factor of the fluid) in energy
due to gyration in magnetic field associated with the fluid and can, in turn,
emit new high energy photons which can leave the fluid and interact in the
external environment, producing new pairs which ``reflect'' a fraction of
energy towards the fluid.

In this work we consider this scenario for a shear flow taking place
in astrophysical jets. We study the mechanism
numerically using nonlinear Large Particle Monte-Carlo Code
developed by Stern et al.\cite{st95}.
 The complete solution of the problem should account for the feedback
of particles on the fluid dynamics and requires detailed numerical treatment
of the hydrodynamical part of the problem. This objective is beyond the scope
of this work and we restrict this first study to a simple model and try to
answer question whether the
supercritical runaway regime exists at  reasonable conditions.

\section{Qualitative consideration }

\begin{figure}
\psfig{figure=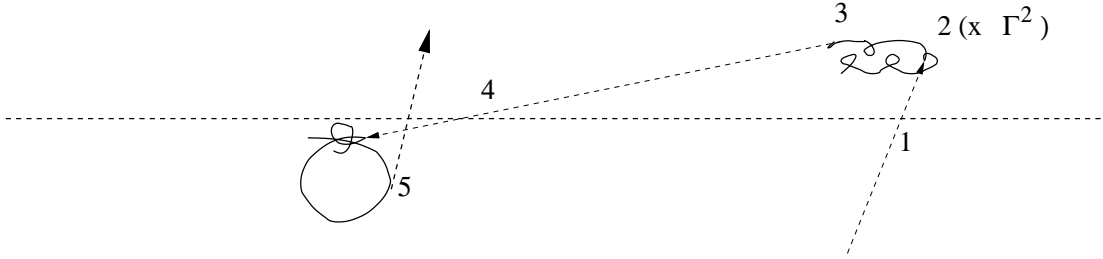,height=3.4cm}
\caption{The scheme of the cascade breeding cycle. The horizontal line separates
 the jet (above, moving to the left) and external environment. See details in the text.
\label{fig:fig1}}
\end{figure}

Very schematically, the process can be split into five steps (see Fig.1).

 Step 1. A high energy external photon (which origin is not important)
enters the jet and interacts with a soft photon producing $e^+ e^-$ pair.

 Step 2. The pair (originally been produced in upstream direction in the
fluid frame) turns around due to gyration in the magnetic field of the jet,
gaining the rest frame energy by factor $\sim \Gamma^2$, where $\Gamma$ is
the Lorentz factor of the jet.

Step 3. The pair Comptonizes soft photons up to high energies.

Step 4. Some of these photons leave the jet and produce pairs in the external
environment.

Step 5. Pairs in the external environment Comptonizes soft photons more or less
isotropically and some of Comptonized photons enter the get again.
This step completes the cycle.

 Each step can be characterized by its ``energy transmission coefficient''
$c_i$ defined as average ratio of total energy of particles with energy
above pair production threshold  before and
after the step.
 $c_2$ is large ($\sim \Gamma^2$), others are smaller than 1.
If $c_1 \times c_2 \times c_3 \times c_4 \times c_5 > 1$ then the
regime is supercritical: each cycle produces more particles than the previous one
and their number grows exponentially. In this case we deal with particle
breeding rather with  particle acceleration: the spectrum of particles
changes slowly (and can evolve to a softer state), but the number and the
total energy of particles rapidly grows.

 Steps 1, 3, 4 and 5 require  a field of soft photons to
provide the conversion of high energy photons into pairs
and production of new high energy photons through inverse Compton
scattering. There are many possible sources of soft photons which were
already considered in the literature:   blackbody and X-ray
radiation of the accretion disk; disk
radiation scattered or reprocessed  in the broad line region \cite{s94,s96,gm96};
external IR radiation
of dust; direct and scattered synchrotron radiation of high energy electrons
in the  jet \cite{gm96}.

 Next condition is the presence of a transversal or chaotic magnetic field,
both in the jet (to provide step 2) and external environment (to provide
isotropization of Comptonized photons at step 5).

\section{Monte-Carlo simulations and their results}

The jet was represented as a cylinder of radius $R_j$  and length $20 R_j$
at distance of $20 R_j$ from the central black hole.
 Actually the jet should be a cone rather than
a cylinder, we neglect the jet divergence for simplicity and adopt constant physical conditions
along the jet. In the course of the simulation the jet undergoes the differential deceleration:
we split the jet into 500 cylindric shells, calculate the momentum transferred to
each shell and decelerate each shell independently from others.  This is a very rough
simplification: the deceleration should depend on $z$ and this will lead to a
complicated situation including formation of internal shocks.

The trajectories of electrons and positrons in the
magnetic field were
simulated directly assuming transversal geometry of the field $H_j$
in the jet and $H_e$ in the external matter.
There is a primary constant soft photon field filling the whole space.

 At the start of simulation the space at $0.9 < r < 1$ and $20 < z < 30$
 (where $r$ and $z$  are the distances from the jet axis  and from
 the black hole in units of $R_j$) is filled
by seed isotropic high energy
photons whose energy density is several orders of magnitude smaller than
the energy density of the jet. After that there is no injection of external photons
and all particles participating in the further simulation are descenders of these
seed photons.

\subsection{Example 1. Weak magnetic field and a ``minimal'' seed radiation}

 In this example we try the simplest variant of the external photon field:
two-component emission of accretion disk. First component is a blackbody spectrum integrated
over disk radius with the maximum temperature 5 eV  and luminosity $L_d \sim 0.5
\cdot 10^{45}$erg s$^{-1}$.
The second component is the power-law with photon index $\alpha = -1.7$ and
cutoff at 50 keV   with luminosity  $L_x \sim 0.5 \cdot 10^{44}$ erg s$^{-1}$.
The jet radius was taken
$R_j = 10^{16}$ cm, the distance from the source $ R = 2 \cdot 10^{17}$ cm.
 The fraction of scattered disk radiation was taken $\xi = 0.05$.
 Magnetic field in
the jet was $H_j= 0.6$ G, which corresponds to the Poynting flux
$10^{43}$ erg s$^{-1}$. The jet energy flux can be much higher, therefore
such magnetic field implies matter dominated jet. The low magnetic field
in this run is not our arbitrary choice since a stronger field reduces
the criticality of the system. At $H_j = 1$G the system is still supercritical,
but the exponential cascade breeding is too slow.
We adopt the total power $L = 10^{45}$ erg s$^{-1}$.

 The total energy release as a function of time is shown in Fig. 2.
 We observe a reasonably fast breeding with e-folding time
$\tau \sim 1.37$ (see Fig. 2). The active layer is rather thin:
a half of the energy release is concentrated within $\delta r \sim 0.02$ from the
jet boundary. At $t \sim 12$ the regime changes: the external
shell decelerates and the active layer gets wider
($\delta r = 0.05$ at $t = 15$ and $\delta r = 0.07$ at $t=20$).
The cascade breeding slows down as the photon path length through the
cycle (see \S 2) increases. The total energy release into photons reached
20\% of the total jet energy at the end of simulation at $t = 20$.
 Our simplified model with a limited number of large particles can not
follow up  the evolution of the system for a longer time correctly.

 While the jet decelerates in our model, the external environment is fixed
at rest. Actually it undergoes a radiative acceleration.
Our assumption is reasonable if the density of external medium is
$n > 10^5$ cm$^{-3}$: then we actually can neglect its acceleration.

 The evolution of the photon spectrum is shown in Fig.~3a. Early spectrum
demonstrate two distinct components: TeV Comptonization peak (mainly
Comptonized emission of the disk UV photons   scattered in the broad-line region)
and a synchrotron maximum.
Qualitatively this spectrum resembles some BL Lac spectra, but in our
simulation the synchrotron maximum is harder than usually observed.

After $t \sim 10$ the spectrum changes: the main peak moves to lower energies
and the synchrotron peak declines. The reason for such evolution is evident:
the system enters the nonlinear stage because soft synchrotron radiation of
the cascade exceeds the initial soft photon field. Comptonization losses
increase while synchrotron losses do not change.


\begin{figure}
\begin{center}
\epsfig{file=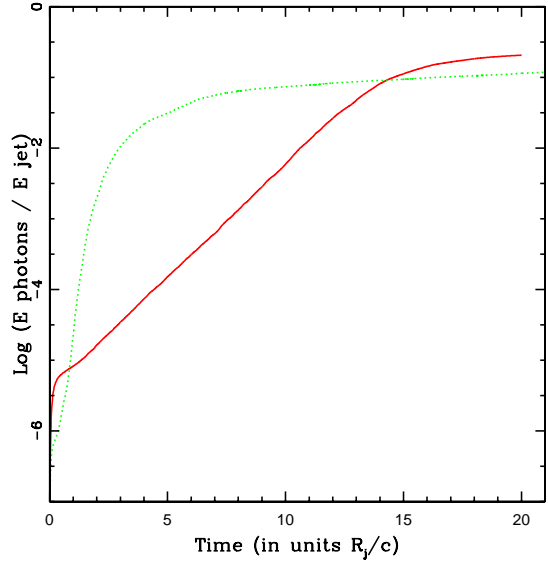,width=9.cm}
\end{center}
\caption{Total fluid energy release into radiation versus time.
Solid (red) curve: Example 1, dotted (green) curve: Example 2 (see text).}
\end{figure}

\begin{figure}
\centerline{\epsfig{file=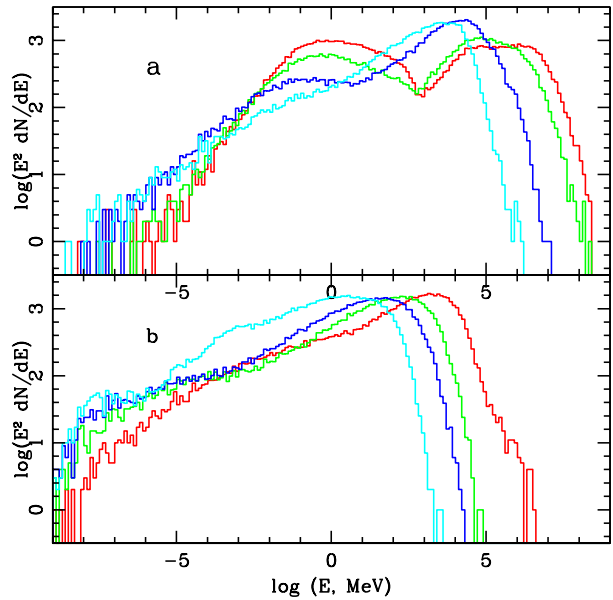,width=9.cm}}
\caption{Evolution of instantaneous spectra of photons. (a) - Example 1
for $t = 5, 12, 15$ and 20: red, green, blue and light blue lines respectively.
(b) Example 2 for $t=2, 4, 6, 15$, the same sequence of colors.
}
\end{figure}

\subsection{Example 2. Strong magnetic field.}

 In this example we take $ R =2 \cdot 10^{16}$ cm, $R_j = 10^{15}$ cm, $H = 100$ G,
$\Gamma = 10$. The Poynting flux at such parameters is $S \sim 10^{45}$erg s$^{-1}$,
which is sufficient for a moderately powerful magnetically dominated jet.
The disk luminosity is $L_d = 10^{45}$erg s$^{-1}$, $L_x = 10^{44}$erg
s$^{-1}$.
The ratio of Compton to synchrotron losses at such parameters is
$U_{\rm rad}/U_{\rm B} \times \Gamma^2 \sim 1$. However, a particle with
comoving Lorentz factor $\gamma > 10^5$ interacts with scattered disk
radiation  in the Klein-Nishina regime and
synchrotron losses dominate. With this reason the system remains subcritical
and our trials have shown no runaway regime at such conditions.

 The situation changes if we add a preexisting synchrotron radiation of the jet
similar to that proposed by Ghisellini \& Madau \cite{gm96}. Then we have more
soft (UV and optical) photons for Comptonization and more X-rays
for conversion of lower energy gamma-rays into pairs. The seed
synchrotron component was taken as a power law spectrum with photon index
$\alpha = - 1.7$   with an exponential
cutoff at 50 keV and the total power $\sim 0.005$ of the jet power.

At the start, the electromagnetic cascade breeds very rapidly (see Fig.2)
with $\tau \sim 0.17$. The time constant is so small due to a short
free path of high energy photons moving in transversal direction as the
soft photon density is much higher than in the previous example. The
active layer is very thin, $\delta r \sim 2 \cdot 10^{-3}$, and the breeding
cycle is short. Evidently, such regime can not last for a long time and
at $t \sim 2$ the active layer decelerates, the photon spectrum gets softer
(see Fig.~3b) and the breeding slows down (Fig. 2). At the end of the
simulation run at $t = 21$ the energy release reached 12\% of the total jet
energy and is probably underestimated (see Discussion).

 The hard to soft evolution of high energy peak of the photon spectrum
shown in Fig.~3 spans almost all range of peak energies observed in blazars.
The latest spectrum peaks in MeV range as in MeV blazars, where the
observed spectra   have a much sharper maximum.
 One also can see a hint on IR--radio synchrotron component observed in
blazars. This component here is less prominent than in blazars, or, in other
terms, intermediate 10 eV - 100 keV luminosity in our simulations has a higher
level than the observed one.

\section{Discussion}

We have demonstrated that the supercritical runaway cascade does develop at
reasonable conditions and can convert into radiation at
least $\sim$ 20\% of the jet kinetic energy. This is certainly not an
ultimate value: with our simplified model we are able to reproduce only
the initial stage of the evolution. Actually we can expect a wealth of
interesting nonlinear phenomena with jet deceleration, formation of internal
shocks, non stationary behaviour producing flares and moving bright features.
A more realistic model should include a detailed treatment of fluid
hydrodynamics coupled with large particle electromagnetic cascade.

 The model can reproduce at least
the high energy component of blazar radiation. On the other hand,
examples presented in this work do  not reproduce the low energy
synchrotron components as prominent as observed in blazars.
The general impression is that our simulated spectra are qualitatively
similar, but flatter and smoother than observed.

The reason why our simulated spectra are flatter (in $E^2 N_{E}$ scale)
and more featureless than the observed spectra is probably a too high
maximal Lorentz factor of pairs produced in the jet, $\gamma_{\max}$.
At early stages, $\gamma_{\max} \sim 10^8$ in Example 1 and
$\gamma_{\max} = 10^6$ in Example 2 (comoving values).
The synchrotron radiation energy is 1000 MeV and
15 MeV (rest frame), respectively,  while the observed synchrotron
peak energy in blazars varies in the range $10^{-6} - 0.1$ MeV.
A high $\gamma_{\max}$ leads to a many generation electromagnetic cascade which
does produce a flat spectrum\cite{sve87}.

 The reason why  $\gamma_{\max}$  is so high in our simulations is a short
scale of the breeding cycle at the early stage: a high energy photon
despite a large opacity can cross the jet boundary, if the latter is sharp.
At later stages, the transition becomes smooth,  the opacity increases
due to generation on new soft photons and highest energy photons cannot
cross a region with large $\Gamma$ increment. Therefore $\gamma_{\max}$
drops and we can see this in Fig.~3. Moreover, in both cases   we can
see the formation of a poorly developed synchrotron component in a right place
in late spectra.
Probably, if we were able to follow up the system evolution for a longer
time and with a more realistic treatment, we would obtain spectra in a better
agreement with observations.

\section*{ACKNOWLEDGMENTS}

The work is supported by the RFBR grant 04-02-16987,
Academy of Finland, Jenny and Antti Wihuri Foundation,
Vilho, Yrj\"o and Kalle V\"ais\"al\"a Foundation, and the NORDITA
Nordic project in High Energy Astrophysics.

\end{document}